\documentclass[sigplan,10pt]{acmart}
\renewcommand\footnotetextcopyrightpermission[1]{}
\usepackage{wrapfig}
\usepackage{mdframed}
\usepackage{listings}
\usepackage{amssymb}

\usepackage{amsmath}
\usepackage{amsthm}
\usepackage{multirow}
\usepackage{hyperref}
\hypersetup{%
  colorlinks=false,
  linkbordercolor=red,
  pdfborderstyle={/S/U/W 1}
}

\usepackage{graphicx}
\usepackage{xspace}

\usepackage{cleveref}
\usepackage{enumitem}

\usepackage[T1]{fontenc}

\usepackage[]{xcolor}

\usepackage{todonotes}

\newcommand{\vc}[1]{\pgwrapper{VC}{#1}} 

\InputIfFileExists{activateeditingmarks}{
}{
    \def\noeditingmarks{}
}

\definecolor{comment-red}{rgb}{0.8,0,0}
\definecolor{dark-green}{rgb}{0.0,0.4,0}
\definecolor{dark-blue}{rgb}{0.0,0.0,0.55}
\definecolor{very-dark-green}{rgb}{0.0,0.3,0}

\ifx\noeditingmarks\undefined

   \definecolor{mygrey}{rgb}{0.7,0.7,0.7}

   \newcommand{\pgwrapper}[2]{\todo[size=\footnotesize]{#1: #2}}

   \InputIfFileExists{activategreybg}{
       \definecolor{comment-red}{rgb}{0.5,0,0}
       \pagecolor{mygrey}
   }{}

  \newcommand{\rnfloat}[1]{\pgwrapper{RRN}{#1}\vspace{-4mm}}  

\else

   \newcommand{\pgwrapper}[2]{}

   \newcommand{\rnfloat}[1]{}  
\fi

\newcommand{\eg}{{e.g.,}}

\newcommand{\sappend}{\texttt{<>}}

\newcommand{\compose}[2]{{#1}$\circ${#2}}
\newcommand{\tof}{\compose{t}{f}}
\newcommand{\fot}{\compose{f}{t}}
\newcommand{\tofone}{\compose{t}{f}1}
\newcommand{\fotone}{\compose{f}{t}1}

\newcommand{\codeurl}{\href{https://github.com/iu-parfunc/verified-instances/tree/da77897ae8b505acb333ab6de224284e5725a53d}{\texttt{http://bit.ly/2qFbei6}}}
\newcommand\LH{Liquid Haskell\xspace}
\newcommand{\il}[1]{\lstinline[language=HaskellUlisses];#1;}
\renewcommand{\eg}{\emph{e.g.}\xspace}

\newcommand{\HaskellStandard}{Haskell 98}

\usepackage{booktabs}   
\usepackage{subcaption} 

\makeatletter\if@ACM@journal\makeatother
\acmJournal{PACMPL}
\acmVolume{1}
\acmNumber{1}
\acmArticle{1}
\acmYear{2017}
\acmMonth{1}
\acmDOI{10.1145/nnnnnnn.nnnnnnn}
\startPage{1}
\else\makeatother
\acmConference{}{}{}
\acmYear{}
\acmISBN{}
\acmDOI{}
\acmPrice{}
\startPage{1}
\fi

\setcopyright{none}             

\usepackage{flushend}
\begin{document}

\title{Deriving Law-Abiding Instances}


\newcommand\NV[1]{{\color{red}{NIKI: #1}}}
\newcommand\RJ[1]{{\color{purple}{RANJIT: #1}}}

\author{Ryan Scott}
\author{Vikraman Choudhury}
\author{Ryan Newton}
\affiliation{
  \institution{Indiana University}            
}
\email{{rgscott,vikraman,rrnewton}@indiana.edu}          

\author{Niki Vazou}
\affiliation{
  \institution{University of Maryland}           
}
\email{nvazou@cs.umd.edu}

\author{Ranjit Jhala}
\affiliation{
  \institution{UC San Diego}           
}
\email{jhala@cs.ucsd.edu}




\ifdefined\withcolor
	 \definecolor{haskellblue}{rgb}{0.0, 0.0, 1.0}
	 \definecolor{haskellstr}{rgb}{0.2, 0.2, 0.6}
	 \definecolor{haskellred}{rgb}{1.0, 0.0, 0.0}
  \definecolor{gray_ulisses}{gray}{0.55}
  \definecolor{castanho_ulisses}{rgb}{0.71,0.33,0.14}
  \definecolor{preto_ulisses}{rgb}{0.41,0.20,0.04}
  \definecolor{green_ulises}{rgb}{0.2,0.75,0}
\else
	\definecolor{haskellblue}{gray}{0.1}
	\definecolor{haskellstr}{gray}{0.1}
	\definecolor{haskellred}{gray}{0.1}
	\definecolor{gray_ulisses}{gray}{0.1}
	\definecolor{castanho_ulisses}{gray}{0.1}
	\definecolor{preto_ulisses}{gray}{0.1}
	\definecolor{green_ulisses}{gray}{0.1}
\fi

%

\definecolor{lcolor}{gray}{0.0}
\definecolor{lappcolor}{gray}{0.0}
\definecolor{lappascolor}{gray}{0.0}

\def\codesize{\small}

\newcommand\showfocus[1]{\color{purple}{\textbf{#1}}}

\lstdefinelanguage{HaskellUlisses} {
    xleftmargin=\parindent,
	basicstyle=\ttfamily\codesize,
	moredelim=[is][\showfocus]{\#}{\#},
	sensitive=true,
	morecomment=[l][\color{gray_ulisses}\ttfamily\itshape\codesize]{--},
	morecomment=[s][\color{gray_ulisses}\ttfamily\itshape\codesize]{\{-}{-\}},
	morestring=[b]",
    escapechar={\`},
	stringstyle=\color{haskellstr},
	showstringspaces=false,
	numberstyle=\codesize,
	numberblanklines=true,
	showspaces=false,
	breaklines=true,
	showtabs=false,
    framexleftmargin=5pt,
    literate={
           {`}{{{$^{\backprime}{}$}}}1
           {?}{{{$\because$}}}1
           {<=}{{$\leq$}}1
           {theta}{{$\theta$}}1
           {rf}{{{\color{lappcolor}f}}}1
           {gf}{{{\color{lappascolor}f}}}1
           {rmap}{{{\color{lappcolor}map}}}3
           {gmap}{{{\color{lappascolor}map}}}3
           {env}{{$\Gamma$}}1
           {|-}{{$\vdash$}}1
           {<=!}{{{\color{lcolor}<=!}}}3
           {!=}{{$\neq$}}1
           {forall}{{$\forall$}}1
           {->}{{$\rightarrow$}}1
           {=*}{{$\eqfun$}}2
           {<=>}{{$\Leftrightarrow$}}3
           {=>}{{$\Rightarrow$}}2
           {<:}{{$\preceq$}}1
           {mempty}{{$\mempty$}}1
           {invalidmappend}{{$\invalidmappend$}}1
           {mappend}{{$\mappend$}}1
           {<>}{{$\mappend$}}1
           {stringMempty}{{$\stringMempty$}}1
           {<+>}{{$\stringMappend$}}1
           {stringMappend}{{$\stringMappend$}}1
           {listMempty}{{[]}}1
           {listMappend}{{++}}2
           {epsilon}{{$\epsilon$}}1
           {eta}{{$\eta$}}1
           {AND}{$\land$.}3
           {&&}{{$\land$}}1
           {_m}{{${}_m$}}1
           {_n}{{${}_n$}}1
           {m^+}{{m${}^{+}$}}2
           },
	emph=
	{[1]
		FilePath,IOError,abs,acos,acosh,all,and,any,appendFile,approxRational,asTypeOf,asin,
		asinh,atan,atan2,atanh,basicIORun,break,catch,ceiling,chr,compare,concat,concatMap,
		const,cos,cosh,curry,cycle,decodeFloat,denominator,digitToInt,div,divMod,drop,
		dropWhile,either,elem,encodeFloat,enumFrom,enumFromThen,enumFromThenTo,enumFromTo,
		error,even,exp,exponent,fail,mapMaybe,filter,flip,floatDigits,floatRadix,floatRange,floor,
		fmap,foldl,foldl1,foldr,foldr1,fromDouble,fromEnum,fromInt,fromInteger,fromIntegral,
		fromRational,fst,gcd,getChar,getContents,getLine,head,id,inRange,index,init,intToDigit,
		interact,ioError,isAlpha,isAlphaNum,isAscii,isControl,isDenormalized,isDigit,isHexDigit,
		isIEEE,isInfinite,isLower,isNaN,isNegativeZero,isOctDigit,isPrint,isSpace,isUpper,iterate,
		last,lcm,length,lex,lexDigits,lexLitChar,lines,log,logBase,lookup,map,mapM,mapM_,max,
		maxBound,posMax,negMax,maximum,maybe,min,minBound,minimum,mod,negate,not,notElem,null,numerator,odd,
		or,ord,pi,pred,primExitWith,print,product,properFraction,putChar,putStr,putStrLn,quot,
		quotRem,range,rangeSize,read,readDec,readFile,readFloat,readHex,readIO,readInt,readList,readLitChar,
		readLn,readOct,readParen,readSigned,reads,readsPrec,realToFrac,recip,rem,repeat,replicate,return,
		reverse,round,scaleFloat,scanl,scanl1,scanr,scanr1,seq,sequence,sequence_,show,showChar,showInt,
		showList,showLitChar,showParen,showSigned,showString,shows,showsPrec,significand,signum,sin,
		sinh,snd,span,splitAt,sqrt,subtract,succ,sum,tail,take,takeWhile,tan,tanh,threadToIOResult,toEnum,
		toInt,toInteger,toLower,toRational,toUpper,truncate,uncurry,undefined,unlines,until,unwords,unzip,
		unzip3,userError,words,writeFile,zip,zip3,zipWith,zipWith3,listArray,doParse,empty,for,initTo,
        assert,compose,checkGE,maxEvens,empty,create,get,set,initialize,idVec,fastFib,fibMemo,
        ex1,ex2,ex3,incr,inc,dec,isPos,positives,find,insert,len,size,union,fromList,initUpto,trim,
        insertSort,decsort,qsort,reverse,append,upperCase, ifM, whileM, get, decrM, diff,
        project, select, leq, elts, keys, dkeys, dfun, addKey, pTrue, emptyRD, rFalse,
                dom, rng, isI, isD, isS, movie1, movie2,  toI, toS, toD, good_titles, runState, ret,
                update, getCtr, setCtr, ctr, rdCtr, wrCtr, ifTest, whileTest, posCtr, zeroCtr, decr, decCtr,
                pread , pwrite , plookup , pcontents, pcreateF , pcreateFP, pcreateD, active, caps, pset, eqP,
                write, contents, alloc, derivP, copyP, createDir, store, copyRec, copySpec,
                forM_, when, flookup, fread, createDir, pcreateFile, isFile, copyFrame, ?
	},
	emphstyle={[1]\color{haskellblue}},
	emph=
	{[2]    Show,Eq,Ord,Num,UpClosed,Comp,Wit,Witness,Meet,Flip,TRUE,Nat,Pos,Neg,IntGE,Plus,List,
        Bool,Char,Double,Either,Float,IO,Integer,Int,Maybe,
        Ordering,Rational,Ratio,ReadS,ShowS,String,Word8,
        InPacket,Tree,Vec,NullTerm,IncrList,DecrList,
        UniqList,BST,MinHeap,MaxHeap,World,RIO,IO,HIO,Post,Pre,
        Privilege, Chain, ChainTy, Range, Dict, RD, Dom, Set, P, Univ, Schema, MovieSchema, RT,
        TDom, TRange, MoviesTable, RTSubEqFlds, RTEqFlds, Disjoint, Union, Ret, Seq, Trans, Map,
        Pure, Then, Else, Exit, Inv, OneState, Priv, Path, FH, Stable,
		Nat
	},
	emphstyle={[2]\color{castanho_ulisses}},
	emph=
	{[3]
		case,class,data,deriving,do,else,if,import,in,infixl,infixr,instance,let,
		module,of,primitive,then,refinement,type,where,forall,bound, newtype,
		measure,reflect,predicate
		, Theorem, Proof, Qed, QED, Definition, match, with , end, Fixpoint, Type,
		Axiom, assume, exists, Lemma, Prop, Inductive, as
	},
	emphstyle={[3]\color{preto_ulisses}\textbf},
	emph=
	{[4]
		quot,rem,div,mod,elem,notElem,seq
	},
	emphstyle={[4]\color{castanho_ulisses}\textbf},
	emph=
	{[5]
		EQ,GT,LT,Left,Right
	},
	emphstyle={[5]\color{preto_ulisses}\textbf},
	emph=
	{[6]
	    axiomatize, measure, inline
	},
	emphstyle={[6]\color{lcolor}}
}

\lstnewenvironment{code}
{\lstset{language=HaskellUlisses,columns=fullflexible,keepspaces,mathescape}}
{}

\lstMakeShortInline[language=HaskellUlisses,mathescape,keepspaces,mathescape,basicstyle=\ttfamily\codesize,breakatwhitespace]@

\newcommand{\makeatcode}{\lstMakeShortInline[style=inline]@}
\newcommand{\makeatchar}{\lstDeleteShortInline@}

\lstdefinelanguage{Pseudo} {
	basicstyle=\ttfamily\codesize,
	sensitive=true,
    mathescape=true,
	morecomment=[l][\color{gray_ulisses}\ttfamily\codesize]{--},
	morecomment=[s][\color{gray_ulisses}\ttfamily\codesize]{\{-}{-\}},
	morestring=[b]",
	showstringspaces=false,
	numberstyle=\codesize,
	numberblanklines=true,
	showspaces=false,
	breaklines=true,
	showtabs=false
}

\begin{abstract}
\LH augments the Haskell language with theorem proving capabilities,
allowing programmers to express and prove class laws.
But many of these proofs require routine, boilerplate code
and do not scale well, as the size of proof terms can grow
superlinearly with the size of the datatypes involved
in the proofs.

We present a technique to derive Haskell proof terms
by leveraging datatype-generic programming. Our observation
is that we can take any algebraic datatype, generate an equivalent
\textit{representation type}, and have \LH automatically construct
(and prove) an isomorphism between the original type
and the representation type. This reduces many proofs down to easy theorems over
simple algebraic ``building block'' types, allowing programmers to write generic
proofs cheaply and cheerfully.
We applied our technique to derive verified instances of
the @Eq@, @Ord@, @Semigroup@, @Monoid@ and @Functor@ Haskell classes
for commonly used algebraic datatypes.

\end{abstract}



\maketitle
\renewcommand{\shortauthors}{R. Scott et al.}

\section{Introduction}

Many widely used type classes abstract over
operators that must obey algebraic laws.
With \LH~\cite{refinement-types-for-haskell},
these type class laws can be encoded as refinement
type specifications.
For instance, @TotalOrd@ extends the Haskell @Ord@ class
with the @total@ method that encodes the proof obligation
that @(<=)@ should be total:
\begin{code}
{-@ class Ord a => TotalOrd a where
     total :: x:a -> y:a -> {x <= y || y <= x} @-}
\end{code}
The type specification of @total@, defined in the
special \LH comments \verb|{-@ ... @-}|, 
states that for all values @x@ and @y@ there exists a proof
that @x <= y@ or @y <= x@, thus encoding the totality of @(<=)@.
Users of @TotalOrd@ can rest assured that @(<=)@ is indeed total,
but when defining an instance of @TotalOrd@, a proof of totality must be provided.

\textit{Haskell} programs can be used to encode such proofs~\cite{refinement-reflection, string-matcher}.
Yet, proof deployment can be tedious.
Implementing many proofs can involve excessive amounts of
boilerplate code. Even worse, the size of some proofs can
grow superlinearly in the size of the data type used,
as the proofs can grow extremely quickly due to the
sheer number of cases one has to exhaust (\S~\ref{sec:scaling-up}).

In this paper, we set out to minimize this boilerplate
and develop a style of proof-carrying programming
that scales well as the size of a data type grows.
To do so, we adapt a style of datatype-generic programming in the tradition of
the Glasgow Haskell Compiler's @GHC.Generics@ module
~\footnote{\url{http://hackage.haskell.org/package/base-4.9.1.0/docs/GHC-Generics.html}}.
That is to say, for some data type about which we want to prove a property, we
first consider a \textit{representation type} which is isomorphic to the
original data type.
This representation type is the composition of several
very small data types. By proving the property in question
for these small, representational data types, we can compose
these proofs and use them to prove the property for the original
data type by taking advantage of the isomorphism between the
original and representation types.

To use @TotalOrd@ as an example of how this would be accomplished, the author
of the @TotalOrd@ class would need to implement
(1) definitions for total orderings on the generic representation types,
and (2) a way to derive a total ordering for a type @a@, reusing a proof
from its representation type (which is provably isomorphic to @a@):
\begin{code}
instance (TotalOrd (Rep a x), GenericIso a)
       => TotalOrd a where
\end{code}

With this generic derivation in hand,
Haskell's standard class resolution will derive the proper (provably correct)
@TotalOrd@ instance for any type that is an instance of @GenericIso@, a class
which carries the proof of isomorphism.
We can automate this process of deriving law-abiding instances further by defining
a Template Haskell function @deriveIso@ which derives the @GenericIso@ instances
with minimal effort.
For instance, one can derive a provably total @Ord@ instance of
the user-defined data type @Nat@ with just:
\begin{code}
data Nat = Zero | Succ Nat
deriveIso ''Nat -- derives: instance GenericIso Nat
instance TotalOrd Nat
\end{code}

%
We provide an implementation of these ideas using
\LH and the Glasgow Haskell Compiler, located at \codeurl.

The contributions of this paper are:

\begin{itemize}

\item We extend Haskell typeclasses to verified typeclasses which have explicit
  proofs of typeclass laws (\S~\ref{sec:motivation}),

\item We propose an extension to GHC generics which adds proofs of isomorphism
  between the original datatype and its representation type, with some machinery
  to automatically derive the proofs (\S~\ref{sec:isos}), and

\item We use the ``generic isomorphism'' machinery to derive verified instances
  for the @Eq@, @Ord@, @Functor@, and @Monoid@ verified typeclasses (\S~\ref{sec:eval}).

\end{itemize}

\section{Law-Abiding Type Classes}\label{sec:motivation}

We start with an overview of our approach for deriving
class instances that are \emph{verified} to satisfy class laws.
First, we briefly review \LH refinement types and show how to formally
\emph{specify} laws as refinement types.
Second, we show how to \emph{manually} create instances that satisfy the laws
(what we call the ``direct approach''), and demonstrate how the direct approach
scales poorly as the size of data types grows.
Third, we show an alternative approach that advocates \emph{composing}
simple verified instances to obtain compound ones.
Then in \cref{sec:isos}, we show how the above process of
composition can be automated via isomorphisms, in the style
of GHC's generic deriving
~\cite{generic-deriving}, yielding an automatic
way of obtaining verified type class instances.

\subsection{\LH as a Theorem Prover}

\LH extends the grammar of Haskell types to include {\em refinements}.  For
example, the following narrows the set of @Int@ values by ruling out zero:

\begin{code}
type NonZero = { n:Integer | n /= 0 }
\end{code}

Refinement types like the above are checked automatically in \LH, which
internally uses an SMT solver.  The \LH implementation assumes that the SMT
solver's notion of integer arithmetic is consistent with Haskell's, and thus
many arithmetic properties become automatically verifiable.

Consider, however, that we want to verify a property such as
@length (tail ls) == length ls - 1@.
Here the @tail@ function is defined with regular Haskell code, and must
somehow be lifted into the refinement logic.  This is the premise of {\em refinement
  reflection}~\cite{refinement-reflection}, a recent addition to \LH.
Using this approach, \LH lifts Haskell definitions into the logic, leaving them
initially uninterpreted, but unfolding their definitions {\em once} every time they
are referenced in an explicit proof of the property.

Thus \LH goes beyond automatically-checked refinements and allows proofs about
Haskell code written as Haskell code.
In these proofs, Haskell's arrow type encodes implication,
Haskell branches encode proof case-splits, and recursion encodes induction.
Together with a library of proof combinators included with \LH,
these enable proofs that are similar to their pencil-and-paper analogues.
We will see examples of such proofs as we proceed in this paper.

\subsection{Specifying Law-Abiding Classes}

\paragraph{Classes}
Recall the following simplified definition of the
@Eq@ and @Ord@ type classes that provide abstractions
for datatypes which support equality and ordering checks:
\begin{code}
class Eq a where
  (==) :: a -> a -> Bool

class Eq a => Ord a where
  (<=) :: a -> a -> Bool
\end{code}

\paragraph{Laws}
Typically, we {require} that any instance
of @Ord@ is a \emph{total order} that
satisfies the following laws:
$$\begin{array}{rl}
\mbox{\emph{Reflexivity }} & \forall x.\ x \leq x \\
\mbox{\emph{Totality    }} & \forall x, y.\ x \leq y \vee y \leq x \\
\mbox{\emph{Antisymmetry}} & \forall x, y.\ x \leq y \wedge y \leq x \Rightarrow x = y \\
\mbox{\emph{Transitivity}} & \forall x, y, z.\ x \leq y \leq z \Rightarrow x \leq z
\end{array}$$


\paragraph{Specifying Laws as Refinement Types}
We can encode the above laws as
\emph{refined function types}:
\begin{code}
type Refl  a = x:a -> {x <= x}
type Total a = x:a -> y:a -> {x<=y || y<=x}
type Anti  a = x:a -> y:a -> {x<=y && y<=x => x == y}
type Trans a = x:a -> y:a -> z:a -> {x<=y<=z => x<=z}
\end{code}
In \LH, these type refinements must be written inside a special comment,
recognized by \LH and separated from the plain Haskell types.  We show only
the \LH type signatures above for brevity.
We write @{p}@ to abbreviate @{v:Proof|p}@ , that
is, the set of values of type @Proof@ such that the predicate @p@ holds.
\footnote{ Here, \il{Proof} is simply a type alias for the unit type \il{()} in
  \LH's library of proof combinators.  Since the proofs carry no useful
  information at runtime, the unit type suffices as a runtime witness to a
  proof.}
Refinement type checking \cite{refinement-reflection} ensures that any
\emph{inhabitant} of @Refl a@ (and respectively, @Total a@, @Trans a@, @Anti a@)
is a concrete \emph{proof} that the corresponding law holds for the type @a@, by
demonstrating that the law holds for all (input) values of type @a@.

\paragraph{Specifying Law-Abiding Classes}
We can specify law-abiding classes by extending the @Ord@ class to a
@VerifiedOrd@ subclass with four more fields that must be inhabited by
\emph{proofs} that demonstrate that the corresponding laws hold for the
instance:
\begin{code}
class Ord a => VerifiedOrd a where
  refl  :: Refl  a
  total :: Total a
  anti  :: Anti  a
  trans :: Trans a
\end{code}

\subsection{Law-Abiding Instances: The Direct Approach}
\label{sec:simple-example}


Next, let's create a @VerifiedOrd@ instance
for a simple data type:
\begin{code}
data A = A Int deriving Eq
instance Ord A where
  (A s1) <= (A s2) = (s1 <= s2)
\end{code}
%
The reflexivity of @A@ can be proved
with proof combinators like so:
\begin{code}
reflA :: Refl A
reflA x@(A s)
  =  x <= x
  =. s <= s
  ** QED
\end{code}
The \emph{implementation} of @reflA@ is a function
that shows that the reflexivity law holds for every
@x :: A@. The function uses the proof combinators
\begin{code}
(=.) :: x:a -> y:{ a | x = y } -> { v:a | v = x }
x =. _ = x

data QED = QED

(**): :: a -> QED -> Proof
_ ** _ = ()
\end{code}
The type of the @(=.)@ function ensures that
the left- and right-hand sides are equal (according
to @(=)@, the SMT solver's notion of equality).
@QED@ and @(**)@ provide a way to link
a chain of equations into a @Proof@. Using these
combinators allows us to build refinement proofs in
``equational reasoning'' style.

Note that the key step for the proof of @reflA@ is the
line @x $\leq$ x@. The underlying SMT solver knows how
to reason about @Int@s directly, so Liquid Haskell is able to
conclude that @x $\leq$ x@ for all @Int@s @x@,
without requiring any lemmas about @Int@ arithmetic.

We can prove antisymmetry, transitivity and totality
for @A@ in much the same way as we did for reflexivity:
\begin{code}
antiA :: Anti A
antiA x@(A s1) y@(A s2)
  =  (x <= y && y <= x)
  =. (s1 <= s2 && s2 <= s1)
  =. (s1 == s2)
  =. (x == y)
  ** QED

transA :: Trans A
transA x@(A s1) y@(A s2) z@(A s3)
  =  (x <= y && y <= z)
  =. (s1 <= s2 && s2 <= s3)
  =. (s1 <= s3)
  =. (x <= z)
  ** QED

totalA :: Total A
totalA x@(A s1) y@(A s2)
  =  (x <= y || y <= x)
  =. (s1 <= s2 || s2 <= s1)
  ** QED
\end{code}

Once these proofs have been established, we can package
them up into a @VerifiedOrd@ instance for @A@:

\begin{code}
instance VerifiedOrd A where
  refl  = reflA
  anti  = antiA
  trans = transA
  total = totalA
\end{code}

\subsection{Scaling Up the Direct Approach}
\label{sec:scaling-up}

Next, let's see how to repeat the process of writing a
@VerifiedOrd@ instance for a more complicated data type.
We shall see that while this is {possible},
the proofs quickly start to become {unpleasant}, as they will require
a lot of boilerplate code.
To see this, consider a data type
with two constructors:

\begin{code}
data B = B1 Int | B2 Int deriving Eq
instance Ord B where
  (B1 s1) <= (B1 s2) = (s1 <= s2)
  (B2 s1) <= (B2 s2) = (s1 <= s2)
  (B1 {}) <= (B2 {}) = True
  (B2 {}) <= (B1 {}) = False
\end{code}

The proof of reflexivity does not change
significantly, as it amounts to adding
another case for the additional constructor:

\begin{code}
reflB :: Refl B
reflB x@(B1 s)
  =  (x <= x)
  =. (s <= s)
  ** QED
reflB x@(B2 s)
  =  (x <= x)
  =. (s <= s)
  ** QED
\end{code}

The proof of antisymmetry, however,
becomes a bit more complicated.
We now require a case for every
pairwise combination of constructors:


\begin{code}
antiB :: Anti B
antiB x@(B1 s1) y@(B1 s2)
  =  (x <= y && y <= x)
  =. (s1 <= s2 && s2 <= s1)
  =. (s2 == s1)
  =. (x == y)
  ** QED
antiB x@(B1 s1) y@(B1 s2)
  =  (x <= y && y <= x)
  =. (s2 <= s1 && s1 <= s2)
  =. (s2 == s1)
  =. (y == x)
  ** QED
antiB x@(B1 {}) y@(B1 {})
  =  (x <= y && y <= x)
  =. (True && False)
  =. False
  ** QED
antiB x@(B1 {}) y@(B1 {})
  =  (x <= y && y <= x)
  =. (False && True)
  =. False
  ** QED
\end{code}

With multiple constructors, there are
cases where the hypothesis does not
hold---namely, when comparing a
@B1@ value with a
@B2@ value.
As the hypothesis reduces to @False@,
the entire implication is vacuously true,
so concluding with @False@ suffices
to prove the output refinement.

\paragraph{Boilerplate Blowup}
However, something worrying has happened here.
The proof of antisymmetry for @A@ only took
two cases, whereas the corresponding proof for
@B@ took four cases.
If we were to add a third constructor, then
the antisymmetry proof would take nine cases.
In other words, the size of this proof is growing
quadratically with the number of constructors!

The other proofs needed for @VerifiedOrd@
also grow quickly.
Like antisymmetry, the proof of totality grows
quadratically, since it must consider every
pairwise combination of two constructors.
The proof of transitivity has an even more noticeable
increase in size growth, since it must match on
every combination of \textit{three} @B@ values:
while the one-constructor variant of
the proof of transitivity has one case, the
two-constructor variant would have eight cases,
and a three-constructor variant would have 27 cases.

Perhaps even more troublesome than the size of
these proofs themselves is the fact that most
of these cases are sheer boilerplate.
For instance, the proof of antisymmetry follows
a predictable pattern. For the cases where the
constructors are both the same, we compare the
fields of the constructors, appeal to properties
of @Int@ arithmetic, and conclude that the two
values are equal.
For the cases where different constructors are
being matched, one comparison will end up being
@False@, causing the whole hypothesis to be
@False@.
This is routine code that is begging to be
automated with a proof-reuse technique.

\section{Deriving Law-Abiding Instances}
\label{sec:isos}

Having seen the tedium of manually constructing proofs, we present
a solution. Notably, our approach does not require adding new features to
\LH itself---instead, we use a technique based on
extensions already found in the Glasgow Haskell Compiler (GHC).

We adapt an approach from the datatype-generic programming literature where we
take an algebraic data type and construct a \textit{representation type}
which is isomorphic to it~\cite{generic-deriving}. The representation type itself
is a composition of small data types which represent primitive notions such as
single constructors, products, sums, and fields. We also establish a type class for
witnessing the isomorphism between a data type and its representation type.

With these tools, we can shift the burden of proof from the original data type
(which may be arbitrarily complex) to the handful of simple data types which
make up representation types. Moreover, since \textit{all} \HaskellStandard{} data types can be
expressed in terms of these representational building blocks, proving a property
for these data types is enough to prove the property for this whole class of algebraic
data types.

\subsection{A Primer on Datatype-Generic Programming}
\label{sec:basic-representational-types}

To build up representation types, we build upon the API from the
@GHC.Generics@ module~\cite{generic-deriving}. First, we utilize a
type class which captures the notion of conversion to and from a
representation type:

\begin{code}
class Generic a where
  type Rep a :: * -> *
  from :: a -> Rep a x
  to :: Rep a x -> a
\end{code}

The @Rep@ type itself will always be some combination of the following
data types:
\footnote{The actual implementation features another data type, \il{M1}, which
is used only for metadata. For the sake of simplicity, we have left it out
of the discussion in this paper.}

\begin{itemize}
\item @data U1 p = U1@. This is used to represent a constructor
       with no fields.
\item @newtype Rec0 c p = Rec0 c@. This is used to represent a single
      field in a constructor.
\item @data (f :*: g) p = (f p) :*: (g p)@. This is used to represent
      the choice between two consecutive \emph{fields} in a constructor.
\item @data (f :+: g) p = L1 (f p) | R1 (g p)@. This is used to
      represent the choice between two consecutive \emph{constructors}
      in a data type.
\end{itemize}

\vc{Also need to mention \il{V1} for empty types}

Recalling the @B@ data type from earlier:

\begin{code}
data B = B1 Int | B2 Int
\end{code}

We define its canonical @Generic@ instance like so:

\begin{code}
instance Generic B where
  type Rep B = Rec0 Int :+: Rec0 Int
  from (B1 i) = L1 (Rec0 i)
  from (B2 i) = R1 (Rec0 i)
  to (L1 (Rec0 i)) = B1 i
  to (R1 (Rec0 i)) = B2 i
\end{code}

Here, we see that because @B@ has two constructors (@B1@ and @B2@), the @(:+:)@
type is used once to represent the choice between @B1@ and @B2@. The @Int@ field
of each constructor is likewise represented with a @Rec0@ type. We call this
instance ``canonical'' because with GHC's @DeriveGeneric@ extension, this
instance is generated automatically with only this line of code:

\begin{code}
deriving instance Generic B
\end{code}

It should be emphasized that the four types @U1@, @Rec0@,
@(:*:)@, and @(:+:)@
are enough to represent \textit{any} \HaskellStandard
~\footnote{They are however not enough to represent the full spectrum of generalized
abstract data types (GADTs) ~\cite{generic-deriving}. Some other generic programming
libraries ~\cite{instant-generics,yakushev} present different designs that allow
representing some features of GADTs, but the question of how to incorporate
GADTs into a \il{GHC.Generics}-style API remains open.}
data type. For instance, if one
were to add more fields to the @B1@ constructor, then its corresponding
@Rep@ type would change by
adding additional occurrences of @(:*:)@ for each field. Therefore,
these four data types conveniently provide a unified way to describe the
structure of any data type, a property which will be useful shortly.

While @Generic@ is convenient for quickly coming up with representation types,
it alone isn't enough for our needs, as we need to be able to use the
\emph{proof} that the @from@ and @to@ functions form an isomorphism. In pursuit
of that goal, we define a subclass of @Generic@ with two proof methods that
express the fact that @from@ and @to@ are mutual inverses.

\begin{code}
class Generic a => GenericIso a where
  `\tof` :: x:a -> { to (from x) == x }
  `\fot` :: x:Rep a x -> { from (to x) == x }
\end{code}

\vc{\il{x:Rep a x} is potentially confusing}

To demonstrate how the proofs in a @GenericIso@ instance look, we give an
example instance for @B@:

\begin{code}
instance GenericIso B where
  `\tof` x@(B1 i)
    =  to (from x) =. to (L1 (Rec0 i))
    =. x ** QED
  `\tof` x@(B2 i)
    =  to (from x) =. to (R1 (Rec0 i))
    =. x ** QED
  `\fot` x@(L1 (Rec0 i))
    =  from (to x) =. from (B1 i)
    =. x ** QED
  `\fot` x@(R1 (Rec0 i))
    =  from (to x) =. from (B2 i)
    =. x ** QED
\end{code}

Unlike @Generic@, there is no built-in GHC mechanism for deriving instances
of @GenericIso@, so one might reasonably worry that @GenericIso@
is itself a source of boilerplate.
We use Template Haskell~\cite{template-haskell}
to mimic GHC's @deriving@ mechanism and automatically derive
@GenericIso@ instances.
Concretely, we define the Template Haskell function @deriveIso@ that, given a
name of a type constructor, derives the declarations
of the corresponding instances of @Generic@ and @GenericIso@.
\begin{code}
deriveIso :: Name -> Q [Dec]
\end{code}

As a demonstration, all of the code
for the @Generic@ and @GenericIso@ instances for @B@ written earlier in this section
can be reduced to:

\begin{code}
data B = B1 Int | B2 Int
deriveIso ''B
\end{code}
where
@''B@ is the Template Haskell @Name@ that represents the type constructor @B@.

%
%
%
%

\subsection{Proofs over Representation Types}
\label{sec:product-example}

Having identified the four basic data types which can be composed in various ways
to form representation types, the next task is to write proofs for these four
types. We will do so by continuing our earlier @VerifiedOrd@ example from
\Cref{sec:motivation}, and in the process show how one can obtain a valid
total ordering for any algebraic data type by using this technique.

The @U1@ data type has an extremely simple @Ord@ instance:

\begin{code}
instance Ord (U1 p) where
  U1 <= U1 = True
\end{code}

The @VerifiedOrd@ instance is similarly straightforward, so we will elide
the details here.

The @Ord@ instance for the @Rec0@ type will look familiar:

\begin{code}
instance Ord c => Ord (Rec0 c p) where
  (Rec0 r1) <= (Rec0 r2) = (r1 <= r2)
\end{code}

This is essentially the same @Ord@ instance that we used for
@A@ in \Cref{sec:simple-example}, except abstracted to an arbitrary
field of type @c@. The @VerifiedOrd@ instance for @Rec0@ also
mirrors that of @A@, so we will also leave out the details here.

The @(:*:)@ type, which serves the role of representing two fields in a
constructor, is also the simplest possible product type, with two conjuncts.  We
can enforce a valid total order on such a type by using the lexicographic
ordering.
\footnote{There are many possible orderings on products, but only lexicographic
  ordering preserves the total order properties.}
We first check if the left fields are equal. If so, we compare the
right fields. Otherwise, we return the comparison on the left fields:

\begin{code}
instance (Ord (f p), Ord (g p)) =>
          Ord ((f :*: g) p) where
  (x1 :*: y1) <= (x2 :*: y2) =
    if x1 == x2 then y1 <= y2 else x1 <= x2
\end{code}

It can be shown that given suitable @VerifiedOrd@ proofs for the fields' types
@f p@ and @g p@, this ordering for @(:*:)@ is reflexive:

\begin{code}
leqProdRefl
  :: (VerifiedOrd (f p), VerifiedOrd (g p))
  => t:((f :*: g) p) -> { t <= t }
leqProdRefl t@(x :*: y) =
     (t <= t)
  =. (if x == x then y <= y else x <= x)
  =. y <= y
  =. True ? refl y
  ** QED
\end{code}

Note that we use an additional
proof combinator @(?)@ here:

\begin{code}
(?) :: (Proof -> a) -> Proof -> a
f ? y = f y
\end{code}

One should read @(?)@ as being ``prove the equational step on the
left-hand side by using the lemma on the right-hand side''. In the case of
@leqProdRefl@, we were able to prove that @y $\leq$ y@ is
true precisely because of the assumption that @y@ was reflexive.
The remaining proofs of antisymmetry, transitivity, and totality for @(:*:)@
can be found in \Cref{product-type-proofs}. Putting
all of these proofs together gives us the following @VerifiedOrd@
instance:

\begin{code}
instance (VerifiedOrd (f p), VerifiedOrd (g p))
       => VerifiedOrd ((f :*: g) p) where
  refl    = leqProdRefl
  antisym = leqProdAntisym
  trans   = leqProdTrans
  total   = leqProdTotal
\end{code}

In a similar vein, we can come up with a @VerifiedOrd@ instance for the @(:+:)@
type. @(:+:)@ not only represents choice between two constructors, it is also
the simplest possible sum type, with two disjuncts. A total ordering on sums is
defined so that everything in the @L1@ constructor is less than everything in
the @R1@ constructor:

\begin{code}
instance (Ord (f p), Ord (g p)) =>
          Ord ((f :+: g) p) where
  (L1 x) <= (L1 y) = x <= y
  (L1 x) <= (R1 y) = True
  (R1 x) <= (L1 y) = False
  (R1 x) <= (R1 y) = x <= y
\end{code}

Here is an example of a @VerifiedOrd@-related proof for @(:+:)@,
establishing reflexivity:

\begin{code}
leqSumRefl
  :: (VerifiedOrd (f p), VerifiedOrd (g p))
  => u:((f :+: g) p) -> { u <= u }
leqSumRefl s@(L1 x) =  (s <= s)
                    =. x <= x
                    =. True ? refl x
                    ** QED
leqSumRefl s@(R1 y) =  (s <= s)
                    =. y <= y
                    =. True ? refl y
                    ** QED
\end{code}

This proof bears a strong resemblance to the reflexivity proof for @B@
in \Cref{sec:simple-example}. This similarity is intended, as the structure of the
@B@ data type is quite similar to that of @(:+:)@. The
remaining proofs for @(:+:)@ can be found in \Cref{sum-type-proofs}).
Finally, we obtain the following
@VerifiedOrd@ instance for @(:+:)@:

\begin{code}
instance (VerifiedOrd (f p), VerifiedOrd (g p))
       => VerifiedOrd ((f :+: g) p) where
  refl    = leqSumRefl
  antisym = leqSumAntisym
  trans   = leqSumTrans
  total   = leqSumTotal
\end{code}

We wish to place particular emphasis on the fact that these @VerifiedOrd@
instances are compositional. That is, we can put together whatever combination
of @(:+:)@, @(:*:)@, @U1@, and @Rec0@ we wish, and
we will ultimately end up with a structure which has a valid @VerifiedOrd@
instance.
This is crucial, as it ensures that this technique scales
up to real-world data types.

\subsection{Reusing Proofs}
\label{sec:iso-example}

Given a @VerifiedOrd@ instance for a representation type, how can we relate
it back to the original data type to which it is isomorphic? The answer lies in
the @GenericIso@ class from before. @GenericIso@ has enough
power to take a @VerifiedOrd@ proof for one type and reuse it for
another type.

To begin, we will need a way to compare two values of a type that is an instance
of @Generic@, given that its representation type @Rep@ is
an instance of @Ord@:

\begin{code}
leqIso :: (Ord (Rep a x), Generic a)
       => (a -> a -> Bool)
leqIso x y = (from x) <= (from y)
\end{code}

We can straightforwardly prove that @leqIso@ is a total order:

\begin{code}
leqIsoRefl
  :: (VerifiedOrd (Rep a x), GenericIso a)
  => x:a -> { leqIso x x }
leqIsoRefl x =  leqIso x x
             =. (from x) <= (from x)
             =. True ? refl (from x)
             ** QED
\end{code}

The proof of antisymmetry relies on the fact that @from@ is an injection, which
follows from the proof of isomorphism.

\begin{code}
fromInj :: GenericIso a => x:a -> y:a
        -> { from x == from y ==> x == y }
fromInj x y =
     from x == from y
  =. to (from x) == to (from y)
  =. x == to (from y) ? `\tof` x
  =. x == y ? `\tof` y
  ** QED

leqIsoAntisym
  :: (VerifiedOrd (Rep a x), GenericIso a)
  => x:a -> y:a
  -> { leqIso x y && leqIso y x => x == y }
leqIsoAntisym x y =
     (leqIso x y && leqIso y x)
  =. ((from x) <= (from y) && (from y) <= (from x))
  =. (from x) == (from y)
      ? antisym (from x) (from y)
  =. x == y ? fromInj x y
  ** QED

leqIsoTrans
  :: (VerifiedOrd (Rep a x), GenericIso a)
  => x:a -> y:a -> z:a
  -> { leqIso x y && leqIso y z => leqIso x z }
leqIsoTrans x y z =
     (leqIso x y && leqIso y z)
  =. ((from x) <= (from y) && (from y) <= (from z))
  =. (from x) <= (from z)
      ? trans (from x) (from y) (from z)
  =. leqIso x z
  ** QED

leqIsoTotal
  :: (VerifiedOrd (Rep a x), GenericIso a)
  => x:a -> y:a
  -> { leqIso x y || leqIso y x }
leqIsoTotal x y =
     (leqIso x y || leqIso y x)
  =. ((from x) <= (from y) || (from y) <= (from x))
  =. True ? total (from x) (from y)
\end{code}

Now we put it all together and write the @VerifiedOrd@ instance that was
begging to be discovered:

\begin{code}
instance (Ord (Rep a x), GenericIso a)
       => Ord a where
  (<=) = leqIso

instance (VerifiedOrd (Rep a x), GenericIso a)
       => VerifiedOrd a where
  refl    = leqIsoRefl
  antisym = leqIsoAntisym
  trans   = leqIsoTrans
  total   = leqIsoTotal
\end{code}

The above two instances take the proofs of
@VerifiedOrd@ for representation types and reuse them
to construct proofs for @any@
isomorphic data type.
More importantly, we can use these instances to define many additional
@VerifiedOrd@ instances with almost no additional effort.

\subsection{Some Complete Examples}

With the above machinery, writing a @VerifiedOrd@ instance becomes a
breeze. We can now rewrite the earlier @VerifiedOrd B@
instance, which was written in the direct approach, and greatly simplify it
using the generic approach:

\begin{code}
data B = B1 Int | B2 Int deriving Eq
deriveIso ''B
instance Ord B
instance VerifiedOrd B
\end{code}

This small amount of code does a tremendous amount of heavy lifting.
Recall (\S~\ref{sec:basic-representational-types}) for @Generic@ and @GenericIso@:

\begin{code}
instance Generic B where
  type Rep B = Rec0 Int :+: Rec0 Int
  ...

instance GenericIso B where ...
\end{code}

Type class resolution will fill in the implementations for the @Ord@ and @VerifiedOrd@
instances for @B@, if we have @Ord@ and @VerifiedOrd@ instances for
@Int@, @Rec0@ and @(:+:)@.
A @VerifiedOrd Int@ instance is trivial
to create, as the SMT solver's reasoning about @Int@s makes the proofs
simple, and we demonstrated how to write the proofs for @(:+:)@ in
\Cref{sec:product-example}.

Our derivation technique, as presented, works for recursive datatypes too.
For instance assume the recursive definition of natural numbers.
\begin{code}
data Nat = Zero | Suc Nat deriving Eq
\end{code}

Then we derive a @VerifiedOrd@ instance for @Nat@
simply by deriving all the appropriate @Generic@, @GenericIso@ and @Ord@
classes\footnote{Note that the methods in the derived instance are only guaranteed to
terminate for strictly positive datatypes.}.

\begin{code}
deriveIso ''Nat
instance Ord Nat
instance VerifiedOrd Nat
\end{code}

%
%
%
%
%
%
%
%
%

\section{Evaluation}\label{sec:eval}

To evaluate our approach for deriving lawful instances,
we extended a set of commonly used Haskell type
classes with associated proof obligations (summarized in Table~\ref{fig:evaluation:classes})
and implemented proof carrying instances for the
Haskell data types of Table~\ref{fig:eval:datatypes}.
Our implementation can be accessed at \codeurl.
%
In this section, we describe the five lawful type classes
(\cref{sec:eval:classes}) and
the law-abiding instances that we derived for them
(\cref{sec:eval:instances}).
We conclude by summarizing the benefits (\cref{sec:comparison})
and limitations (Sections~\ref{sec:limitations} and \ref{sec:performance}) of our technique.

\begin{table}[t]
\begin{code}
class Eq a => VerifiedEq a where
  refl  :: ReflEq  a
  sym   :: SymEq  a
  trans :: TransEq a

class Ord a => VerifiedOrd a where
  refl  :: Refl  a
  total :: Total a
  anti  :: Anti  a
  trans :: Trans a

class Semigroup a => VerifiedSemigroup a where
  assoc :: Assoc a

class Monoid a => VerifiedMonoid a where
  lident :: LIdent a
  rident :: RIdent a

class Functor f => VerifiedFunctor f where
  fmapId      :: FmapId f
  fmapCompose :: FmapCompose f
\end{code}
\caption{Summary of the law-abiding type classes.}
\label{fig:evaluation:classes}
\end{table}

\begin{table}[t]
\begin{code}
data Identity a   = Identity a
data Maybe a      = Nothing | Just a
data Either a b   = L a | R b
data List a       = Nil | Cons a (List a)
data Triple a b c = MkTriple a b c
\end{code}
\caption{Summary of the evaluated data-types.}
\label{fig:eval:datatypes}
\end{table}

\subsection{Lawful Type Classes}\label{sec:eval:classes}

We used refinement types to specify the laws for five
standard type classes as presented in~\cref{fig:evaluation:classes}.

\paragraph{1. Total Orders}
Our primary example from \cref{sec:motivation}
was the @Ord@ type class, which can be
verified to be a total order.

\paragraph{2. Equivalences}
Next we specify the equivalence properties in @Ord@'s superclass, @Eq@.
\begin{code}
class Eq a where
  (==) :: a -> a -> Bool
\end{code}
Equality should be an \emph{equivalence} relation---that is, it should satisfy
the laws of reflexivity, symmetry, and transitivity (expressed directly as
refined function types):
\begin{code}
type ReflEq  a = x:a -> {x == x}
type SymEq   a = x:a -> y:a -> {x == y => y == x}
type TransEq a = x:a -> y:a -> z:a
               -> {x == y && y == z => x == z}
\end{code}

These type signatures are used in the class methods of @VerifiedEq@ in
Table~\ref{fig:evaluation:classes}.
The process for generically creating @VerifiedEq@
instances is extremely similar to the
process for @VerifiedOrd@, as outlined
in \cref{sec:motivation}.

\paragraph{3. Semigroups}
Next, we specify the associativity law for semigroups.
%
The @Semigroup@ class comes equipped with a binary operation @($\sappend$)@
that provides a way to combine two values into one.
\begin{code}
class Semigroup a where
  ($\sappend$) :: a -> a -> a
\end{code}
The proof obligation for ($\sappend$) is that it is associative:
\begin{code}
type Assoc a = x:a -> y:a -> z:a
             -> {x $\sappend$ (y $\sappend$ z) = (x $\sappend$ y) $\sappend$ z}
\end{code}

The process of generically creating @VerifiedSemigroup@ instances slightly differs from
that of @VerifiedOrd@ (from \cref{sec:motivation}), since @Semigroup@ features a
class method with the type parameter in the result position of a function---that
is, the type parameter is used covariantly as well as contravariantly.
This means that in order to turn a
@VerifiedSemigroup a@ instance to
a @VerifiedSemigroup b@ instance
with @GenericIso@, one must use
the @to@ function---which was unused up to this point---as well as @from@.
%

\paragraph{4. Monoids}
On top of @Semigroup@, its subclass @Monoid@
\footnote{At the time of writing, \il{Monoid} is not actually a subclass of
  \il{Semigroup} in GHC's \il{base} library.
For the sake of making the presentation more
convenient, however, we will pretend it is.}
grants the ability to conjure up an identity element:
\begin{code}
class Semigroup a => Monoid a where
  empty :: a
\end{code}

@Monoid@ has two more proof obligations which dictate how @empty@ should
interact with the @($\sappend$)@ operation. @empty@ acts as the left and right
identity element:
\begin{code}
type LIdent a = x:a -> { empty $\sappend$ x = x }
type RIdent a = x:a -> { x $\sappend$ empty = x }
\end{code}

There is an interesting question to be asked about whether one can sensibly
write generic @Semigroup@ or @Monoid@ instances for sum types.
Unlike the @Eq@ or @Ord@ classes, where it is straightforward to implement
generic instances for types with multiple constructors (represented by the type
@(:+:)@), for @Semigroup@ and @Monoid@ the choice is not clear.
Trying to combine values from different constructors with @($\sappend$)@ would
require arbitrarily picking whether the left or right constructor should be
used, for instance.
As a result, we did not pursue any @VerifiedSemigroup@ or @VerifiedMonoid@
instances for sum types.


\paragraph{5. Functors}
Finally we specify the laws on the @Functor@ class:
\begin{code}
class Functor (f :: * -> *) where
  fmap :: (a -> b) -> f a -> f b
\end{code}
We use the standard Haskell definitions for identity and composition:
\begin{code}
id :: a -> a
id z = z

(.) :: (b -> c) -> (a -> b) -> a -> c
(.) f g x = f (g x)
\end{code}
to specify that functors preserve identity and composition:
\begin{code}
type FmapId f
  = z:(f a) -> {fmap id z = z}
type FmapCompose f
  = x:(b -> c) -> y:(a -> b) -> z:(f a)
  -> {fmap (x . y) z = (fmap x . fmap y) z}
\end{code}

Unlike the previous four classes that are defined over types (of kind @(*)@),
@Functor@ is defined over type constructors (of kind @(* -> *)@).
To derive law-abiding instances over these kinds of classes,
we need to generalize our earlier machinery to work over @(* -> *)@-kinded types.

\paragraph{Generic Derivations for Type Constructors.}
The @Generic1@ class handles @(* -> *)@-kinded types.
\begin{code}
class Generic1 (f :: * -> *) where
  type Rep1 f :: * -> *
  from1 :: forall a. f a -> Rep1 f a
  to1   :: forall a. Rep1 f a -> f a
\end{code}

The @GenericIso1@ class extends @Generic1@, expressing that @to1@ and @from1@
form a natural isomorphism.

\begin{code}
class Generic1 f => GenericIso1 (f :: * -> *) where
  `\tofone` :: forall a. x:f a -> { to1 (from1 x) == x }
  `\fotone` :: forall a. x:Rep1 a x -> { from1 (to1 x) == x }
\end{code}

Next, it is necessary to increase our set of representational data
types slightly, since implementing @Functor@ demands that we ask more
interesting questions about the structure of data types. To see why that is the
case, observe this data type's @Functor@ instance:
\begin{code}
newtype Phantom a = Phantom Int
instance Functor Phantom where
  fmap f (Phantom i) = Phantom i
\end{code}

This is different than the @Functor@ instance for
this very similar data type:
\begin{code}
newtype Identity a = Identity a
instance Functor Identity where
  fmap f (Identity x) = Identity (f x)
\end{code}

The only distinction between the internal structure
of @Phantom@ and @Identity@ is that @Identity@'s
field is an occurrence of its type parameter.
In order to query this property generically,
we need additional data types that mark
occurrences of the type parameter:
\begin{code}
newtype Par1 p = Par1 p
newtype Rec1 f p = Rec1 (f p)
newtype (f :.: g) p = Comp1 (f (g p))
\end{code}

These three types are used in conjunction
with @Generic1@ exclusively. To see how they are used,
here is a sample @Generic1@ instance:
\begin{code}
data T a = MkT Int a (Maybe a) [[a]]
instance Generic1 T where
  type Rep1 T =
    Rec0 Int :*: Par1 :*: Rec1 Maybe
             :*: ([] :.: Rec1 [])
  from1 (T a1 a2 a3 a4) =
    Rec0 a1 :*: Par1 a2 :*: Rec1 a3
            :*: Comp1 (fmap Rec1 a4)
  to1 (Rec0 a1 :*: Par1 a2 :*: Rec1 a3
               :*: Comp1 a4) =
    T a1 a2 a3 (fmap (\(Rec1 x) -> x) a4)
\end{code}
We see that @Par1@ handles direct occurrences
of the type parameter, @Rec1@ handles cases
where the type parameter is underneath an application
of some type, and @(:.:)@ is used when there
are multiple levels of type applications covering
the type parameter. For all other field types,
@Rec0@ is used.

Finally, following~\cref{sec:motivation}, we define the Template Haskell
derivation function @deriveIso1@ that, given the name of a data type constructor,
derives the proper @Generic1@ and @GenericIso1@ instances.

\begin{code}
deriveIso1 :: Name -> Q [Dec]
\end{code}

\subsection{Law-Abiding Instances}\label{sec:eval:instances}

We used our approach to derive law-abiding instances of the above type classes
for data types of @Identity@, @Maybe@, @Either@, @List@, and @Triple@ as defined
in \cref{fig:eval:datatypes}.
As discussed in~\cref{sec:eval:classes}, we do not attempt to derive @Semigroup@
and @Monoid@ instances for the sum types @Maybe@, @Either@, and @List@.
We selected the five data types in \cref{fig:eval:datatypes} because
they provide a healthy variety of structure,
encompassing types with products, sums and nullary
constructors. Moreover, they provide interesting
test cases for @VerifiedFunctor@ as, \eg, the @List@
type features the type parameter @a@ in both a direct
occurrence and underneath the @List@ type constructor
(in the @Cons@ constructor).

To recap the advantage of our approach, we describe
how each instance was verified, using the @VerifiedFunctor@
instance for @List@ as an example.

\paragraph{At the library site,} the developer defines
the verified class together with its laws:
%
%
\begin{code}
type FmapId f = forall a. z:(f a) -> {fmap id z = z}
type FmapCompose f
  = forall a b c. x:(b -> c) -> y:(a -> b) -> z:(f a)
  -> {fmap (x . y) z = (fmap x . fmap y) z}

class Functor f => VerifiedFunctor f where
  fmapId      :: FmapId f
  fmapCompose :: FmapCompose f
\end{code}
\vc{This discussion seems arbitrary at this point}
To allow semi-automatic derivation of law-abiding instances,
the library developer needs to provide two further
pieces of code:
\begin{enumerate}
\item the verified instances for the representation types
    needed to support the original data type, and
\item a way to convert a verified instance for the
    representation type back to the original data type.
\end{enumerate}

\paragraph{Code 1.}
In our example,
the library-writer must create \sloppy @VerifiedFunctor@
instances for the @U1@, @Par1@, @Par1@, @(:+:)@,
and @Rec1@ types.
%
%
These instances will be used to derive the @VerifiedFunctor@ instance for @List@
since it has the following representation type:
\begin{code}
type Rep1 List = U1 :+: (Par1 :*: Rec1 List)
\end{code}

\paragraph{Code 2.} Then, one needs to define how to convert a @VerifiedFunctor@ instance
for the representation type of @f@ into a \sloppy @VerifiedFunctor@ instance for
@f@ itself.
\begin{code}
instance (VerifiedFunctor (Rep1 f), GenericIso1 f)
    => VerifiedFunctor f
\end{code}
This instance definition can be defined
using the techniques from \Cref{sec:product-example}.

\paragraph{At the user site,} first the data type is defined.
For our example, we use @List@s.

\begin{code}
data List a = Nil | Cons a (List a)
\end{code}

Next, we use Template Haskell to automate the creation of
@Generic1@ and @GenericIso1@ instances for
the data type:
\begin{code}
deriveIso1 ''List
\end{code}

Finally, we derive the law-abiding instance definition
of @List@ as a @VerifiedFunctor@ by simply
by writing the following instance declaration:
\begin{code}
instance VerifiedFunctor List
\end{code}

\subsection{Proof Burden for Direct and Derived Instances}
\label{sec:comparison}
We would like to emphasize the differences between
our \emph{generic} derivation approach and the direct approach of
writing out the proofs \emph{directly}.

In the direct approach, the library writer does not need to write
anything that resembles Code 2, since there are
no data type conversions to be found.
In this sense, there is a cost to the generic approach
that is not present in the direct approach.
Importantly, though, this cost only has to be paid once
for each class, because this code for converting
@VerifiedFunctor@ instances between types can be
reused for every subsequent data type that needs
a @VerifiedFunctor@ instance.

Additionally, the direct approach's costs significantly
outweigh the generic approach's costs.
To implement Code 1 in the generic approach, one must
write proof code for a certain number of ``building block''
data types, \textit{but no more than that.}
After these proofs have been written, there are no
additional costs that arise later when writing other
verified instances, as these proofs can be reused for
other datatypes that have representation types with
the same underlying building block types.
In contrast, the direct approach requires writing
(and re-writing) proof code for {every} verified
instance.

\subsection{Limitations and Future Work}\label{sec:limitations}

Our current prototype differs from the presentation
in \Cref{sec:isos} in a couple of ways.

\paragraph{\LH Doesn't Support Type Classes}
First, \LH does not fully support refining
all features of type classes of the time of writing.
This is a limitation which could be overcome with a
future implementation. We work around this in our
prototype by using an explicit dictionary style
that is equivalent to how type classes
are desugared internally in GHC.
For instance, we reify the @Eq@ type class as
\begin{code}
data Eq a = Eq { (==) :: a -> a -> Bool }
\end{code}
We then explicitly pass around @Eq@ ``instances''
as data type values. This makes the implementation
a bit more verbose, but is otherwise functionally
equivalent to our presentation earlier in the paper.

\paragraph{Template Haskell Doesn't Support Comments}
The other limitation which our prototype must work
around is the lack of Template Haskell support
for generating comments.
Recall that \LH refinements are expressed in
comments of the form \verb|{-@ ... @-}|.
This poses a challenge for us, as we use
Template Haskell to implement the @deriveIso@
function, which is intended to create @GenericIso@
instances and the associated refinement-containing
comments that accompany the instances.
That is, ideally
\begin{code}
data Foo = Foo
deriveIso ''Foo
\end{code}
would suffice to generate the following Haskell code:
\begin{code}
instance Generic Foo where
  to = ...
  from = ...
instance GenericIso Foo where
  {-@ `\tof` :: x:Foo -> {to (from x) == x}       @-}
  `\tof` = ...
  {-@ `\fot` :: x:Rep Foo x -> {from (to x) == x} @-}
  `\fot` = ...
\end{code}

Unfortunately, Template Haskell currently does
not support splicing in declarations that contain
comments as in the code above, so doing everything
in one fell swoop is not possible at the moment.
To work around this limitation, we require users
to write the comments themselves:
\begin{code}
data Foo = Foo
deriveIso ''Foo
{-@ `\tof` :: x:Foo       -> { to (from x) == x } @-}
{-@ `\fot` :: x:Rep Foo x -> { from (to x) == x } @-}
\end{code}
We intend to resolve this by extending
Template Haskell to support comment
generation.

\subsection{A Note on Performance}\label{sec:performance}

One limitation to watch out for is the efficiency of
the verified instances at runtime. A consequence of
using @GHC.Generics@ is that there are many
intermediate data types used, and this can lead
to runtime performance overheads if GHC does not
optimize away the conversions to and from the
intermediate types.
It is sometimes possible to tune GHC's optimization
flags to achieve performance that is comparable to
direct, hand-written code~\cite{optimizing-generics-is-easy},
but as a general rule, code written with @GHC.Generics@
tends to be slower overall.

We do not offer a solution to this problem in this
paper, but it is worth noting that many of the classes
that we discuss can be derived in GHC through
other means. For instance, one can derive efficient
implementations of the @Eq@, @Ord@, and @Functor@
classes by writing
\begin{code}
data Pair a = MkPair a a
  deriving (Eq, Ord, Functor)
\end{code}

One thing we wish to explore in the future is verifying
instances derived in this fashion.
This will be non-trivial as the code that
GHC derives often uses primitive operations
that can be tricky to reason about.
If this were implemented, we could quickly
verify a set of commonly used type classes
and have them be fast, too.


\section{Aside: Logic}

The idea of proof reuse is motivated from model theory in mathematical
logic. First-order model theory studies properties of models of first-order
theories using tools from universal algebra. In particular, preservation
theorems study the closure properties of classes of models across algebraic
operations. By interpreting Haskell type classes and verified type classes as
algebraic objects, we can borrow these ideas to do generic proving and verified
programming.

A Haskell type class can be interpreted as a signature in the sense of universal
algebra, that is, a collection of function and relation symbols with fixed
arities. Relations are identified with propositions, that is, functions whose
codomain is @Bool@. For example, the type class @Eq@ corresponds
to the signature $\sigma_{Eq}:=(=)$, and the @Ord@ class corresponds to
the signature $\sigma_{Ord}:=(\leq,=)$. ``Type class laws'', expressed as
first-order axioms using refinement reflection are identified as a first-order
theory, that is, a set of first-order statements (identified upto logical
equivalence). For example, for @VerifiedOrd@, we have the theory of total
orders given by $T_{Ord}$ with the axioms for reflexivity, antisymmetry,
transitivity, and totality.

We can now interpret building an instance of a verified type class
model-theoretically. A type is an instance of a verified type class, if it forms
a structure in that signature, and is also a model of the first-order
theory. For example, a type $a$ is an instance of @VerifiedOrd@, if there
are operations $=^a$, $\leq^a$ so that $A:=(a,=^a,\leq^a)$ is a $\sigma_{Ord}$
structure, and $A \vDash T_{Ord}$, that is, $A$ is a model of $T_{Ord}$.

Given a first-order theory $T$ and $K$, the class of models of $T$, one can ask
if $K$ is closed under algebraic operations like products ($P(K)$), coproducts
($C(K)$), substructures ($S(K)$), homomorphic images ($H(K)$), isomorphic images
($I(K)$). The answers to some of these are well known~\cite{hodges1997shorter}.

\begin{itemize}
\item $I(K) = K$ for any $T$.
\item (\emph{\L{}o\'{s}-Tarski}) $S(K) = K$ iff $T$ is universal.
\item $SP(K) = K$ iff $T$ is a Horn-clause theory.
\item (\emph{Birkhoff}) $HSP(K) = K$ iff $T$ is equational.
\end{itemize}

This gives a firm theoretical foundation for our technique for shorter
refinement reflection proofs. The fact that classes of models are closed under
products means that if we can prove a property for two types, then we can
immediately conclude that the property holds for a constructor with those two
types as fields. Similarly, closure under coproducts lets us conclude that if a
property holds for two constructors, then that property holds for a sum type
composed of those two constructors. Closure under substructures means that we
can use an injective embedding to reduce the proof to one for a different
datatype. Lastly, closure under isomorphism lets us say that if we can prove a
property for one data type, then we can conclude the property for any other data
type with an isomorphic structure.

\section{Related Work}

Several languages with dependent types offer some degree of automation via
datatype-generic programming. Dagand~\cite{a-cosmology-of-datatypes} develops
a dependent type theory in Agda which, by encoding inductive data types in a
universe of \textit{descriptions},
allows deriving decidable (and boolean) equality in a straightforward manner.
Al-Sibahi~\cite{the-practical-guide-to-levitation} presents a similar
implementation of described types in Idris, based off of the dependent type theory
by Chapman et al. ~\cite{gentle-art-of-levitation},
and demonstrates its utility in deriving instances of decidable equality,
@Functor@, pretty-printing, and generic traversals.
Altenkirch et al. also develop several universes of types in Epigram, which
can be used to implement generic zipper options
~\cite{altenkirch-2006}.

\LH takes a somewhat different approach to equational reasoning than Agda and Idris.
With refinement reflection, the programmer states the propositions as refinements,
and \LH is tasked with finding the proofs (with some gentle assistance by the
programmer). The proof code simply acts as a guide to the SMT solver
in determining satisfiability. In Agda and Idris, however, more responsibility
is placed on the programmer to implement the details of proofs, as their typecheckers
do not leverage a solver. In this way, refinement reflection inverts the
relative importances of propositions and proofs, and by incorporating
statements from propositions into the SMT solver, \LH makes propositions
``whole-program''.

One thing to note is that while the datatype generic programming techniques in
dependently typed languages like Agda, Idris, and Epigram are strictly more powerful, as
they need to support a richer universe of datatypes than what Haskell offers,
it comes with a burden of a higher learning curve. For instance, Al Sibahi notes
that in the generic programming library he developed for Idris, ``it requires
considerable effort to understand the type signatures for even simple operations.''
~\cite{the-practical-guide-to-levitation} In contrast, the generic programming
library we use here is designed to be relatively straightforward to implement,
simple to explain, and give decently understandable type error messages.

The notion of reusing proofs over isomorphic types is also a familiar idea
in the dependent types community. Barthe and Pons \cite{type-isomorphisms}
formalize a theory of \textit{type isomorphisms} in a modified version of the
Calculus of Inductive Constructions. Type isomorphisms are extremely similar
to the @GenericIso@ class in \Cref{sec:basic-representational-types}.
A type isomorphism between types $A$ and $B$
is essentially a pair of two well typed functions
$f : A \rightarrow B$ and $g : B \rightarrow A$ that are mutual inverses
(i.e, that $f \; (g \; x) = x$ and $g \; (f \; x) = x$ for all $x$) which allow one
to take a proof of a property over $A$ and reuse it for $B$, and vice versa.
Barthe and Pons use as motivation the ability to, for instance, reuse a proof of
Peano (unary) natural numbers, which can be easier to reason about,
for binary natural numbers, which can be used for more efficient algorithms.
The technique could be adapted for inductive data types and their corresponding
representations as well.

Isomorphisms (or equivalences) are also well studied in Homotopy Type Theory,
and having a computational interpretation for univalence would mean that all
type constructors act functorially on isomorphims. This allows one to rewrite
terms between isomorphic types, witnessed by a path, which facilitates
type-generic programming. Some possible applications to generic programming are
discussed by Licata and Harper in their work on 2-dimensional type
theory~\cite{licata2012canonicity}.

\section{Conclusion}
We presented how law-abiding type class instances can be derived via generic programming.
Class laws are encoded as refinement type specifications.
The library author's only responsibility is to provide proofs of the laws on
generic representation types,
and to implement a way to derive a verified instance for a type by reusing
the proofs from its (provably isomorphic) representation type.
Then, Haskell's standard class resolution will derive
provably law-abiding instances.
We used this technique on the commonly used Haskell classes
@Eq@, @Ord@, @Semigroup@, @Monoid@ and @Functor@.
Even though our technique currently suffers from various engineering limitations,
it suggests a clean route towards semi-automated verification of class proofs
by combining datatype-generic programming and type class resolution.


{
\bibliography{misc_bibliography/refs,refs}
}

\clearpage
\appendix
\section{Appendix}

\subsection{Full \il{VerifiedOrd} instance for \il{(:*:)}}
\label{product-type-proofs}

\begin{code}
instance (Ord (f p), Ord (g p)) =>
          Ord ((f :*: g) p) where
  (x1 :*: y1) <= (x2 :*: y2) =
    if x1 == x2 then y1 <= y2 else x1 <= x2

leqProdRefl
  :: (VerifiedOrd (f p), VerifiedOrd (g p))
  => Refl ((f :*: g) p)
leqProdRefl t@(x :*: y) =
     (t <= t)
  =. (if x == x then y <= y else x <= x)
  =. y <= y
  =. True ? refl y
  ** QED

leqProdAntisym
  :: (VerifiedOrd (f p), VerifiedOrd (g p))
  => Anti ((f :*: g) p)
leqProdAntisym p@(x1 :*: y1) q@(x2 :*: y2) =
     (p <= q && q <= p)
  =. ((if x1 == x2 then y1 <= y2 else x1 <= x2) &&
      (if x2 == x1 then y2 <= y1 else x2 <= x1))
  =. (if x1 == x2
        then (y1 <= y2 && y2 <= y1)
        else (x1 <= x2 && x2 <= x1))
  =. (if x1 == x2
        then y1 == y2
        else x1 <= x2 && x2 <= x1) ? antisym y1 y2
  =. (if x1 == x2
        then y1 == y2
        else x1 == x2) ? antisym x1 x2
  =. (x1 == x2 && y1 == y2)
  =. (p == q)
  ** QED

leqProdTrans
  :: (VerifiedOrd (f p), VerifiedOrd (g p))
  => Trans ((f :*: g) p)
leqProdTrans p@(x1 :*: y1) q@(x2 :*: y2) r@(x3 :*: y3) =
  case x1 == x2 of
    True  -> case x2 == x3 of
      True  -> (p <= q && q <= r)
            =. (y1 <= y2 && y2 <= y3)
            =. y1 <= y3 ? trans y1 y2 y3
            =. (if x1 == x3
                  then y1 <= y3
                  else x1 <= x3)
            =. (p <= r)
            ** QED
      False -> (p <= q && q <= r)
            =. (y1 <= y2 && x2 <= x3)
            =. x1 <= x3
            =. (if x1 == x3
                  then y1 <= y3
                  else x1 <= x3)
            =. (p <= r)
            ** QED
    False -> case x2 == x3 of
      True  -> (p <= q && q <= r)
            =. (x1 <= x2 && y2 <= y3)
            =. x1 <= x3
            =. (if x1 == x3
                  then y1 <= y3
                  else x1 <= x3)
            =. (p <= r)
            ** QED
      False -> case x1 == x3 of
        True  -> (p <= q && q <= r)
              =. (x1 <= x2 && x2 <= x3)
              =. (x1 <= x2 && x2 <= x1)
              =. (x1 == x2) ? antisym x1 x2
              =. y1 <= y3
              =. (if x1 == x3
                    then y1 <= y3
                    else x1 <= x3)
              ** QED
        False -> (p <= q && q <= r)
              =. (x1 <= x2 && x2 <= x3)
              =. x1 <= x3 ? trans x1 x2 x3
              =. (if x1 == x3
                    then y1 <= y3
                    else x1 <= x3)
              =. (p <= r)
              ** QED

leqProdTotal
  :: (VerifiedOrd (f p), VerifiedOrd (g p))
  => Total ((f :*: g) p)
leqProdTotal p@(x1 :*: y1) q@(x2 :*: y2) =
     (p <= q || q <= p)
  =. ((if x1 == x2 then y1 <= y2 else x1 <= x2) ||
      (if x2 == x1 then y2 <= y1 else x2 <= x1))
  =. (if x1 == x2
        then (y1 <= y2 || y2 <= y1)
        else (x1 <= x2 || x2 <= x1))
  =. (if x1 == x2
        then True
        else (x1 <= x2 || x2 <= x1)) ? total y1 y2
  =. (if x1 == x2
        then True
        else True) ? total x1 x2
  =. True
  ** QED

instance (VerifiedOrd (f p), VerifiedOrd (g p))
       => VerifiedOrd ((f :*: g) p) where
  refl    = leqProdRefl
  antisym = leqProdAntisym
  trans   = leqProdTrans
  total   = leqProdTotal
\end{code}

\subsection{Full \il{VerifiedOrd} instance for \il{(:+:)}}
\label{sum-type-proofs}

\begin{code}
instance (Ord (f p), Ord (g p)) =>
          Ord ((f :+: g) p) where
  (L1 x) <= (L1 y) = x <= y
  (L1 x) <= (R1 y) = True
  (R1 x) <= (L1 y) = False
  (R1 x) <= (R1 y) = x <= y

leqSumRefl
  :: (VerifiedOrd (f p), VerifiedOrd (g p))
  => Refl ((f :+: g) p)
leqSumRefl s@(L1 x) =  (s <= s)
                    =. x <= x
                    =. True ? refl x
                    ** QED
leqSumRefl s@(R1 y) =  (s <= s)
                    =. y <= y
                    =. True ? refl y
                    ** QED

leqSumAntisym
  :: (VerifiedOrd (f p), VerifiedOrd (g p))
  => Anti ((f :+: g) p)
leqSumAntisym p@(L1 x) q@(L1 y) =
     (p <= q && q <= p)
  =. (x <= y && y <= x)
  =. x == y ? antisym x y
  ** QED
leqSumAntisym p@(L1 x) q@(R1 y) =
     (p <= q && q <= p)
  =. (True && False)
  =. False
  =. p == q
  ** QED
leqSumAntisym p@(R1 x) q@(L1 y) =
     (p <= q && q <= p)
  =. (False && True)
  =. False
  =. p == q
  ** QED
leqSumAntisym p@(R1 x) q@(R1 y) =
     (p <= q && q <= p)
  =. (x <= y && y <= x)
  =. x == y ? antisym x y
  ** QED

leqSumTrans
  :: (VerifiedOrd (f p), VerifiedOrd (g p))
  => Trans ((f :+: g) p)
leqSumTrans p@(L1 x) q@(L1 y) r@(L1 z) =
     (p <= q && q <= r)
  =. (x <= y && y <= z)
  =. x <= z ? trans x y z
  =. (p <= r)
  ** QED
leqSumTrans p@(L1 x) q@(L1 y) r@(R1 z) =
     (p <= q && q <= r)
  =. (x <= y && True)
  =. (p <= r)
  ** QED
leqSumTrans p@(L1 x) q@(R1 y) r@(L1 z) =
     (p <= q && q <= r)
  =. (True && False)
  =. (p <= r)
  ** QED
leqSumTrans p@(L1 x) q@(R1 y) r@(R1 z) =
     (p <= q && q <= r)
  =. (True && y <= z)
  =. (p <= r)
  ** QED
leqSumTrans p@(R1 x) q@(L1 y) r@(L1 z) =
     (p <= q && q <= r)
  =. (False && y <= z)
  =. (p <= r)
  ** QED
leqSumTrans p@(R1 x) q@(L1 y) r@(R1 z) =
     (p <= q && q <= r)
  =. (False && True)
  =. (p <= r)
  ** QED
leqSumTrans p@(R1 x) q@(R1 y) r@(L1 z) =
     (p <= q && q <= r)
  =. (x <= y && False)
  =. (p <= r)
  ** QED
leqSumTrans p@(R1 x) q@(R1 y) r@(R1 z) =
     (p <= q && q <= r)
  =. (x <= y && y <= z)
  =. x <= z ? trans x y z
  =. (p <= r)
  ** QED

leqSumTotal
  :: (VerifiedOrd (f p), VerifiedOrd (g p))
  => Total ((f :+: g) p)
leqSumTotal p@(L1 x) q@(L1 y) =
     (p <= q || q <= p)
  =. (x <= y || y <= x)
  =. True ? total x y
  ** QED
leqSumTotal p@(L1 x) q@(R1 y) =
     (p <= q || q <= p)
  =. (True || False)
  ** QED
leqSumTotal p@(R1 x) q@(L1 y) =
     (p <= q || q <= p)
  =. (False || True)
  ** QED
leqSumTotal p@(R1 x) q@(R1 y) =
     (p <= q || q <= p)
  =. (x <= y || y <= x)
  =. True ? total x y
  ** QED

instance (VerifiedOrd (f p), VerifiedOrd (g p))
       => VerifiedOrd ((f :+: g) p) where
  refl    = leqSumRefl
  antisym = leqSumAntisym
  trans   = leqSumTrans
  total   = leqSumTotal
\end{code}


\end{document}